\documentstyle[epsfig,12pt,a4p]{article}

\newcommand{\pipipipi}{\mbox{$\pi^+\pi^-\pi^+\pi^-$ }}

\begin{document}
\begin{titlepage}
\def\footnoterule{\hrule width 1.0\columnwidth}
\begin{tabbing}
put this on the right hand corner using tabbing so it looks
 and neat and in \= \kill
\> {9 July 1997}
\end{tabbing}
\bigskip
\bigskip
\begin{center}{\Large  {\bf A study of the
centrally produced \pipipipi channel
in pp interactions at 450 GeV/c}
}\end{center}
\bigskip
\bigskip
\begin{center}{        The WA102 Collaboration
}\end{center}\bigskip
\begin{center}{
D.\thinspace Barberis$^{  5}$,
W.\thinspace Beusch$^{   5}$,
F.G.\thinspace Binon$^{   7}$,
A.M.\thinspace Blick$^{   6}$,
F.E.\thinspace Close$^{  4}$,
K.M.\thinspace Danielsen$^{ 12}$,
A.V.\thinspace Dolgopolov$^{  6}$,
S.V.\thinspace Donskov$^{  6}$,
B.C.\thinspace Earl$^{  4}$,
D.\thinspace Evans$^{  4}$,
B.R.\thinspace French$^{  5}$,
T.\thinspace Hino$^{ 13}$,
S.\thinspace Inaba$^{   9}$,
A.V.\thinspace Inyakin$^{  6}$,
T.\thinspace Ishida$^{   9}$,
A.\thinspace Jacholkowski$^{   5}$,
T.\thinspace Jacobsen$^{  12}$,
G.V.\thinspace Khaustov$^{  6}$,
T.\thinspace Kinashi$^{  11}$,
J.B.\thinspace Kinson$^{   4}$,
A.\thinspace Kirk$^{   4}$,
W.\thinspace Klempt$^{  5}$,
V.\thinspace Kolosov$^{  6}$,
A.A.\thinspace Kondashov$^{  6}$,
A.A.\thinspace Lednev$^{  6}$,
V.\thinspace Lenti$^{  5}$,
S.\thinspace Maljukov$^{   8}$,
P.\thinspace Martinengo$^{   5}$,
I.\thinspace Minashvili$^{   8}$,
K.\thinspace Myklebost$^{   3}$,
T.\thinspace Nakagawa$^{  13}$,
K.L.\thinspace Norman$^{   4}$,
J.M.\thinspace Olsen$^{   3}$,
J.P.\thinspace Peigneux$^{  1}$,
S.A.\thinspace Polovnikov$^{  6}$,
V.A.\thinspace Polyakov$^{  6}$,
Yu.D.\thinspace Prokoshkin$^{\dag  6}$,
V.\thinspace Romanovsky$^{   8}$,
H.\thinspace Rotscheidt$^{   5}$,
V.\thinspace Rumyantsev$^{   8}$,
N.\thinspace Russakovich$^{   8}$,
V.D.\thinspace Samoylenko$^{  6}$,
A.\thinspace Semenov$^{   8}$,
M.\thinspace Sen\'{e}$^{   5}$,
R.\thinspace Sen\'{e}$^{   5}$,
P.M.\thinspace Shagin$^{  6}$,
H.\thinspace Shimizu$^{ 14}$,
A.V.\thinspace Singovsky$^{  6}$,
A.\thinspace Sobol$^{   6}$,
A.\thinspace Solovjev$^{   8}$,
M.\thinspace Stassinaki$^{   2}$,
J.P.\thinspace Stroot$^{  7}$,
V.P.\thinspace Sugonyaev$^{  6}$,
K.\thinspace Takamatsu$^{ 10}$,
G.\thinspace Tchlatchidze$^{   8}$,
T.\thinspace Tsuru$^{   9}$,
G.\thinspace Vassiliadis$^{\dag   2}$,
M.\thinspace Venables$^{  4}$,
O.\thinspace Villalobos Baillie$^{   4}$,
M.F.\thinspace Votruba$^{   4}$,
Y.\thinspace Yasu$^{   9}$.
}\end{center}

\begin{center}{\bf {{\bf Abstract}}}\end{center}

{
The reaction
$ pp \rightarrow p_{f} (\pi^{+}\pi^{-}\pi^{+}\pi^{-}) p_{s}$
has been studied at 450 GeV/c
in an experiment designed to search for gluonic states.
A spin analysis has been performed and the $dP_T$ filter applied.
In addition to the well known
$f_{1}(1285)$ there is evidence for two $J^{PC}$~=~$2^{-+}$ states
called the $\eta_2(1620)$ and $\eta_2(1875)$ and a broad scalar
called the $f_0(2000)$.
The production of these states as a function
of the $dP_T$ kinematical filter
shows the behaviour expected for $q \overline q$ states.
In contrast,
there is evidence for two states at 1.45 GeV and at 1.9 GeV which do not show
the behaviour observed for $q \overline q$ states.
}
\bigskip
\bigskip
\begin{center}
{\bf Dedicated to the memory of our colleague and friend Yuri Prokoshkin}
\end{center}
\bigskip
\bigskip\begin{center}{{Submitted to Physics Letters}}
\end{center}
\bigskip
\bigskip
\begin{tabbing}
aba \=   \kill
$^\dag$ \> \small
Deceased. \\
$^1$ \> \small
LAPP-IN2P3, Annecy, France. \\
$^2$ \> \small
Athens University, Nuclear Physics Department, Athens, Greece. \\
$^3$ \> \small
Bergen University, Bergen, Norway. \\
$^4$ \> \small
School of Physics and Astronomy, University of Birmingham, Birmingham, U.K. \\
$^5$ \> \small
CERN - European Organization for Nuclear Research, Geneva, Switzerland. \\
$^6$ \> \small
IHEP, Protvino, Russia. \\
$^7$ \> \small
IISN, Belgium. \\
$^8$ \> \small
JINR, Dubna, Russia. \\
$^9$ \> \small
High Energy Accelerator Research Organization (KEK), Tsukuba, Ibaraki 305,
Japan. \\
$^{10}$ \> \small
Faculty of Engineering, Miyazaki University, Miyazaki, Japan. \\
$^{11}$ \> \small
RCNP, Osaka University, Osaka, Japan. \\
$^{12}$ \> \small
Oslo University, Oslo, Norway. \\
$^{13}$ \> \small
Faculty of Science, Tohoku University, Aoba-ku, Sendai 980, Japan. \\
$^{14}$ \> \small
Faculty of Science, Yamagata University, Yamagata 990, Japan. \\
\end{tabbing}
\end{titlepage}
\setcounter{page}{2}
\bigskip
\par
The WA76 and WA91 collaborations
have studied the centrally produced
$\pi^{+}\pi^{-}\pi^{+}\pi^{-}$ final state in the reaction
\begin{equation}
pp \rightarrow p_{f} (\pi^{+}\pi^{-}\pi^{+}\pi^{-}) p_{s}
\label{eq:a}
\end{equation}
at 85~\cite{re:f},
300~\cite{re:a},
and 450 GeV/c \cite{re:wa914pi}.
The subscripts f and s indicate the
fastest and slowest particles in the laboratory respectively.
In addition to the $f_1(1285)$, which was observed at all energies, the
WA76 collaboration observed two new states at 1.45 and 1.9 GeV in their
300~GeV/c data~\cite{re:a}.
In contrast,
no clear evidence was seen for these states in the 85~GeV/c
data of the same experiment~\cite{re:f}.
The increase in cross section with increased incident energy~\cite{re:a}
is consistent with the formation of these states via a
double Pomeron exchange mechanism, which is predicted
to be a source of
gluonic states \cite{re:b}.
\par
The WA91 experiment~\cite{re:wa914pi}, which studied reaction (1)
at 450 GeV/c,
confirmed the existence of these states and,
from a spin analysis, showed that the peak at 1.45 GeV had
$I^{G}J^{PC}=0^{+}0^{++}$
and that the peak at 1.9 GeV had
$I^{G}J^{PC}=0^{+}2^{++}$; hence they were called
the $f_{0}(1450)$ and $f_{2}(1900)$ respectively.
In a reanalysis of this data~\cite{WAINT},
the WA91 collaboration showed that it was possible
to describe the $f_0$(1450) as being due to the interference between the
$f_0(1300)$ and the $f_0(1500)$.
\par
This paper presents new results from the WA102 experiment
which is a continuation of the WA76 and WA91 programme and
has more than a factor of five
increase in statistics.
This enables a detailed spin analysis to be performed
and also a study as a function of $dP_T$, which is the difference
in the transverse momentum vectors of the two exchanged particles~\cite{WADPT}
and has been proposed as a glueball-$q \overline q$ filter~\cite{ck97}.
\par
The data come from experiment WA102
which has been performed using the CERN Omega Spectrometer.
The layout of the Omega Spectrometer used in this run is similar to that
described in ref.~\cite{wa9192} with the replacement of the
OLGA calorimeter by GAMS~4000~\cite{gams}.
\par
Reaction~(\ref{eq:a})
has been isolated from the sample of events having six outgoing
tracks by first imposing the following cuts on the components of the
missing momentum:
$|$missing~$P_{x}| <  14.0$ GeV/c,
$|$missing~$P_{y}| <  0.12$ GeV/c and
$|$missing~$P_{z}| <  0.08$ GeV/c,
where the x axis is along the beam
direction.
A correlation between
pulse-height and momentum
obtained from a system of
scintillation counters was used to ensure that the slow
particle was a proton.
\par
The quantity $\Delta$, defined as
$ \Delta = MM^{2}(p_{f}p_{s}) - M^{2}(\pi^{+}\pi^{-}\pi^{+}\pi^{-})$,
was then calculated for each event and
a cut of $|\Delta|$ $\leq$ 3.0 (GeV)$^{2}$ was used to select the
$\pi^{+}\pi^{-}\pi^{+}\pi^{-}$
channel. Events containing a fast $\Delta^{++}(1232) $
were removed if $M(p_{f} \pi^{+}) < 1.3 $ GeV, which left
1~167~089 centrally produced events.
\par
Fig.~\ref{fi:a}a shows the acceptance corrected
$\pi^{+}\pi^{-}\pi^{+}\pi^{-}$
effective mass spectrum renormalised to the total number of observed events.
The mass spectrum is very similar to that observed by
experiments WA76 and WA91, namely, a clear peak at 1.28~GeV associated with the
$f_1$(1285), a peak at 1.45~GeV called the $f_0$(1450),
which is possibly
due to an interference effect between the $f_0(1300)$ and $f_0(1500)$, and
a broad enhancement at 1.9 GeV called the $f_2(1900)$.
The mass spectrum
has been fitted
using three Breit-Wigners
representing
the $f_{1}(1285)$, the 1.45 GeV peak and the $f_{2}(1900)$
plus a background of the form
$a(m-m_{th})^{b}exp(-cm-dm^{2})$, where
$m$ is the
$\pi^{+}\pi^{-}\pi^{+}\pi^{-}$
mass,
$m_{th}$ is the
$\pi^{+}\pi^{-}\pi^{+}\pi^{-}$
threshold mass and
a, b, c, d are fit
parameters. Reflections from the $\eta\pi^{+}\pi^{-}$
decay of the
$\eta^{\prime}$ and $f_{1}(1285)$
give small enhancements in the
$\pi^{+}\pi^{-}\pi^{+}\pi^{-}$
mass spectrum in the 0.8 and 1.1~GeV regions
due to a
slow $\pi^{0}$ from the decay of an $\eta$
falling within the missing momentum cuts.
In order to get a correct description of the
$\pi^{+}\pi^{-}\pi^{+}\pi^{-}$
mass spectrum two histograms representing a Monte Carlo
simulation of each reflection have been included in the fit.
The Breit-Wigner describing the $f_1$(1285) has been convoluted with a
Gaussian with $\sigma$~=~12~MeV representing the experimental resolution
in this mass region.
The masses and widths determined from the
fit
are
\begin{tabbing}
abaabddb \= mdsd \= mm \=1234 pm 23mm\= mmmyyyy  \= gam \= mm \= 312 pm12 \=gev
 \kill
\> $M_1$ \>= \>1281 $\pm$  1 \> MeV \>$\Gamma_1$  \>= \>24 $\pm$ 3 \>MeV \\
\> $M_2$ \>= \>1445 $\pm$ 4 \> MeV \>$\Gamma_2$  \>= \>95 $\pm$ 15 \>MeV \\
\> $M_3$ \>= \>1920 $\pm$ 20 \> MeV \>$\Gamma_3$  \>= \>450 $\pm$ 60 \>MeV.
\end{tabbing}
\par
Close and Kirk~\cite{ck97}
have proposed that when the centrally produced system is analysed
as a function of the parameter $dP_T$, which is the difference
in the transverse momentum vectors of the two exchange particles~\cite{WADPT},
states with large (small) internal angular momentum will be enhanced at
large (small) $dP_T$.
A study of the \pipipipi mass spectrum as a function of $dP_T$ is presented in
fig.~\ref{fi:a}b), c) and d)
for $dP_T$~$\leq$~0.2~GeV,
0.2~$\leq$~$dP_T$~$\leq$~0.5~GeV and
$dP_T$~$\geq$~0.5~GeV respectively.
A dramatic effect is observed; the $f_1$(1285) signal has virtually disappeared
at low $dP_T$ whereas the $f_0(1450)$ and $f_2(1900)$ signals remain.
\par
A fit to
figs.~\ref{fi:a}b), c) and d)  has been performed (not shown) using the
resonance parameters fixed from the fit to the total mass spectrum.
A poor quality fit results which indicates that the underlying resonance
structure may be more complicated.
In order to try to understand  the resonances that are present in the system
an extraction of the partial waves is required.
\par
A spin-parity analysis of the
$\pi^{+}\pi^{-}\pi^{+}\pi^{-}$
channel
has been performed using an isobar model~\cite{re:wa914pi}.
Assuming that
only angular momenta up to 2 contribute,
the intermediate
states considered are
\begin{tabbing}
abddb \= mmmmmmmmm \= mmmmmmmmm \= mmmmmmmmm \= mmmmmmmmm  \= mmmm\kill
\> $\sigma\sigma $, \> $\sigma(\pi^{+}\pi^{-})_{S wave} $,\>
$\sigma(\pi^{+}\pi^{-})_{P wave} $, \> $\sigma(\pi^{+}\pi^{-})_{D wave} $, \>
$\sigma \rho^{0} $, \\
\>$\rho^{0}\rho^{0} $, \> $\rho^{0}(\pi^{+}\pi^{-})_{S wave} $, \>
$\rho^{0}(\pi^{+}\pi^{-})_{P wave} $, \> $\rho^{0}(\pi^{+}\pi^{-})_{D wave} $,
\\
\> $a_{1}(1260)\pi $, \> $a_{2}(1320)\pi $,
\> $f_{2}(1270)\sigma $, \> $f_{2}(1270)(\pi^{+}\pi^{-})_{S wave} $, \\
\> $f_{2}(1270)\rho^0 $, \> $f_{2}(1270)f_2(1270) $  \\
\end{tabbing}
where $\sigma$ stands for the low mass $\pi\pi$ S-wave amplitude squared.
Two parameterisations of the $\sigma$ have been tried, that of
Au, Morgan and Pennington~\cite{re:AMP} and that of Zou and
Bugg~\cite{re:zbugg}.
The amplitudes have been calculated in the spin-orbit (LS)
scheme using spherical
harmonics.
\par
In order to perform a spin parity analysis the
log likelihood function, ${\cal L}_{j}=\sum_{i}\log P_{j}(i)$,
is defined by combining the probabilities of all events in 40 MeV
$\pi^{+}\pi^{-}\pi^{+}\pi^{-}$ mass bins from 1.02 to 2.82 GeV.
The incoherent sum of various
event fractions $a_{j}$ is calculated
so as to include more than one wave in the fit,
\begin{equation}
{\cal L}=\sum_{i}\log \left(\sum_{j}a_{j}P_{j}(i) +
(1-\sum_{j}a_{j})\right)
\end{equation}
where the term
$(1-\sum_{j}a_{j})$ represents the phase space background.
The negative log likelihood function ($-{\cal L} $) is then minimised using
MINUIT~\cite{re:MINUIT}. Coherence between different $J^{P}$ states
and between different isobar amplitudes of a given $J^{P}$ have
been neglected in the fit.
Different combinations of waves and isobars have been tried and
insignificant contributions have been removed from the final fit.
As was already mentioned
in the WA91 publication~\cite{re:wa914pi},
in the analysis of the $f_{1}(1285)$ and the $f_0(1450)$ peak there is
little change
in the result if the $\rho\rho$ amplitude is used
instead of
the $\rho(\pi\pi)_{P wave}$.
Using Monte Carlo simulations it has been found that the feed-through
from one spin parity to another is negligible and that the
peaks in the spin analysis can not be produced by phase space
or acceptance effects.
\par
The results of the best fit are shown in
fig.~\ref{fi:b} for the total data sample and in
fig.~\ref{fi:c} for three ranges of $dP_T$.
The $f_{1}(1285) $ is clearly seen in the
$J^{P}=1^{+}$~$\rho\rho $.
Superimposed on the
$J^{P}=1^{+}$~$\rho\rho $ wave for the total sample is
the Breit Wigner convoluted with a Gaussian used to describe the
$f_1$(1285) in the fit to the mass spectrum.
As can be seen the $f_1$(1285) is well represented in both size and shape.
{}From fig.~\ref{fi:c} it can be seen that the $f_1$(1285) dies away
for $dP_T$~$\leq$~0.2~GeV.
\par
The $J^{P} =0^{+} \rho\rho $ distribution
in fig.~\ref{fi:b} shows a peak at 1.45~GeV
together with a broad enhancement around 2~GeV.
In order to test if the peak at 1.45~GeV is due to
$\sigma \sigma$ rather than $\rho \rho$
the change in log likelihood
in the four 40 MeV bins around 1.45~GeV
has been calculated
by replacing
the
$J^{P} =0^{+} \rho\rho $ amplitude by
the
$J^{P} =0^{+} \sigma\sigma $ amplitude. The likelihood
decreases by
$\Delta{\cal L} = 9000$  corresponding to
$n=\sqrt{2\Delta{\cal L}}=134$ standard deviations.
This raises a possible problem in interpreting
the peak at 1.45 GeV with the $f_0(1500)$
since the analysis of the $4\pi$ channel in $p \overline p$ by
Crystal Barrel~\cite{cb4pi} and in radiative
$J/\psi$ decay~\cite{jpsi4pi} shows that  the $f_0(1500)$ decays via
$\sigma \sigma$. A possible explanation could be that the
interference between the $f_0(1300)$ and $f_0(1500)$ produces the
observed $\rho\rho$ decay. This question should be answered in the future
by a study of centrally produced $4\pi^0$ events.
\par
{}From fig.~\ref{fi:c} it can be seen that the peak
in the $J^{P} =0^{+} \rho\rho $ wave
around 1.45 GeV remains
for $dP_T$~$\leq$~0.2~GeV while the
$J^{P} =0^{+} $
enhancement at 2.0~GeV becomes less
important: which shows that the $dP_T$ effect is not simply a $J^P$ filter.
A fit has been performed to the
$J^{P} =0^{+} \rho\rho $ amplitude in fig.~\ref{fi:b} in the 1.45~GeV region
using a single Breit-Wigner and gives a mass of
1445~$\pm$~4~MeV and a width of 200~$\pm$~20~MeV. The mass is compatible
with the fit to the total mass spectrum
although the width is broader; it is nonetheless
compatible with a fit performed to the mass spectrum for
$dP_T$~$\leq$~0.2~GeV.
A similar fit has been performed to the
$J^{P} =0^{+} \rho\rho $ distributions shown in fig.~\ref{fi:c}.
The mass and width found in each interval are constant within the errors
of the fit. This would indicate that either the peak is due
to a single resonance or that if it is due to interference
between the $f_0(1300)$ and $f_0(1500)$ then both states have a
similar $dP_T$ dependence.
\par
In order to test the interference hypothesis
a fit has first been performed to the total
$J^{P} =0^{+} \rho\rho $
distribution in fig.~\ref{fi:b} using a K matrix formalism~\cite{KMATRIX}
including poles to describe the interference between the
$f_0(1300)$, the $f_0(1500)$ and a possible state at 2~GeV.
The result of the fit is superimposed on the
$J^{P} =0^{+} \rho\rho $ distribution shown in fig.~\ref{fi:b} and well
describes the data. The resulting resonance parameters are given in
table~\ref{ta:a}. The parameters of the $f_0(1300)$ and
$f_0(1500)$ are very similar to those found by Crystal Barrel~\cite{CBNEW}.
\par
As can be seen from
fig.~\ref{fi:b}
neither the
$J^{P} =2^{+}$ $a_{2}(1320)\pi $ nor the
$J^{P} =2^{+}$ $f_{2}(1270)(\pi\pi)_{S wave} $ alone
can describe the $f_{2}(1900)$ peak observed in the mass spectrum.
However,
the sum of the two waves
accounts for most of the
$f_{2}(1900)$ signal.
The $a_{2}(1320)\pi$ distribution and the
$f_{2}(1270)(\pi\pi)_{S wave}$ distribution peak at different masses
which suggests that the $f_{2}(1900)$ could be composed of
two $J^{PC}=2^{++}$ resonances
but neither distribution is well represented by a Breit-Wigner.
However,
the sum of the
$J^{P} =2^{+}$ $a_{2}(1320)\pi $ and
$J^{P} =2^{+}$ $f_{2}(1270)(\pi\pi)_{S wave} $ waves has been fitted with
a single Breit-Wigner, giving a better description, with
M = 1960 $\pm$ 30 MeV and $\Gamma$ = 460 $\pm$ 40 MeV.
These values are compatible
with those coming from the fit to the total mass spectrum.
\par
In addition to these waves, an extra wave,
not required in the WA76 and WA91 analyses,
is found to be necessary in the
fit, namely the
$J^{P} =2^{-}$ $a_{2}(1320)\pi $ wave. The addition of this wave
increases the log likelihood by 2700.
The resulting
$J^{P} =2^{-}$ $a_{2}(1320)\pi $ wave is shown in
fig.~\ref{fi:b} where a broad enhancement is observed which peaks at 1.6 GeV.
Other experiments have observed evidence for $J^{PC}$~=~$2^{-+}$ states
in this region.
In the reaction $\gamma \gamma \rightarrow \eta \pi \pi$ the
Crystal Ball collaboration~\cite{cbgg} reported a
$J^{PC}$~=~$2^{-+}$ state decaying to
$a_{2}(1320)\pi $ and $a_0(980) \pi$
with a mass of 1881~$\pm$~32~~$\pm$~40~MeV and a width of
221~$\pm$~92~$\pm$~44~MeV. In $p \overline p$ interactions the
Crystal Barrel collaboration~\cite{cbetapipi} reported the observation of two
$J^{PC}$~=~$2^{-+}$ states in the $\eta \pi \pi$ final state.
The first state,
called the $\eta_2(1620)$,
had a mass of 1645~$\pm$~14~$\pm$~5~MeV
and a width of 180~$\pm$~40~$\pm$~25~MeV and
the second state,
called the $\eta_2(1875)$,
had a mass
of 1875~$\pm$~20~$\pm$~35 MeV
and a width of 250~$\pm$~25~$\pm$~45~MeV.
\par
As can be seen from fig.~\ref{fi:b} the
$J^{P} =2^{-}$ $a_{2}(1320)\pi $ wave observed in this experiment is consistent
with the two $\eta_2$ resonances discussed above with both states decaying
to $a_2(1320) \pi $. The masses and widths found from the fit are
given in table~\ref{ta:a}.
\par
In order to calculate the contribution of each resonance as a function
of the $dP_T$ the distributions in fig.~\ref{fi:c} have been fitted with
the parameters of the resonances fixed to those obtained from the
fits to the total data. The results of the fits are given in table~\ref{ta:b}.
\par
As is observed from figs.~\ref{fi:a} and \ref{fi:c} and from table~\ref{ta:b}
the $f_1(1285)$ signal almost disappears at
small $dP_T$. In addition, the $2^{-+} a_2(1320)\pi$ signal is also
suppressed at small $dP_T$. This behaviour is
consistent with the signals being due to
standard $q \overline q$ states~\cite{ck97}.
\par
The $0^{++} \rho \rho$ wave shows that
the $f_0(1450)$ peak is still prominent at small $dP_T$.
If the peak is interpreted as the interference between the $f_0$(1300)
and $f_0(1500)$ then, as can be seen from table~\ref{ta:b},
both must have a similar $dP_T$ dependence.
The broad peak at 2.0 GeV in the $0^{++} \rho \rho$ wave,
called the $f_0(2000)$, is less prominent
at small $dP_T$ and is more consistent with other $q \overline q$ states.
\par
Both the
$J^{P} =2^{+}$ $a_{2}(1320)\pi $ and
$J^{P} =2^{+}$ $f_{2}(1270)(\pi\pi)_{S wave} $ distributions have a very
similar
$dP_T$ behaviour consistent with there being just one resonance at 1.9 GeV
with two decay modes.
In fig.~\ref{fi:c} the sum of these two waves is presented. As can be
seen from this figure and
from table~\ref{ta:b}
at small $dP_T$ the $f_2(1900)$ signal is still important.
This is the first evidence of a non-zero spin resonance produced
at small $dP_T$ and hence shows that the $dP_T$ effect is not just
a $J^P$~=~$0^+$ filter.
\par
It is interesting to note that the prominence of the $f_0$(1450) and
$f_2(1900)$ signals in fig.~\ref{fi:a}b) for $dP_T$~$\leq$~0.2~GeV is not only
due to the fact that the $f_0(1450)$ and $f_2(1900)$ signals
are strong at small $dP_T$ but also due to the fact that the
$\eta_2$ signals are suppressed.
\par
In summary, in
the centrally produced
$\pi^{+}\pi^{-}\pi^{+}\pi^{-}$ mass spectrum two interesting structures
are observed
at 1.45 and 1.9~GeV.
A spin analysis shows that the underlying resonance structure
is more complex. The peak at 1.45~GeV is found to have
$J^{PC}=0^{++}$ and to decay to $\rho\rho$; it
can either be described as
a single Breit-Wigner
or as an interference between the $f_0(1300)$ and $f_0(1500)$.
The peak observed at 1.9~GeV
appears as a broad enhancement in the mass spectrum and is found to
decay to $a_{2}(1320)\pi$
and $f_{2}(1270)\pi\pi$ with $J^{PC}=2^{++}$. Due to the
difficulty in describing the individual
$a_{2}(1320)\pi$
and $f_{2}(1270)\pi\pi$
contributions by
a Breit Wigner it is likely that this is one state, called the $f_2(1900)$,
with two decay modes.
Confirmation is found for two $J^{PC}=2^{-+}$ resonances,
called the $\eta_2(1620)$ and $\eta_2(1875)$
decaying to
$a_{2}(1320)\pi$.
There is also evidence for a broad state with $J^{PC}=0^{++}$ decaying
to $\rho \rho$, called the $f_0(2000)$.

\newpage
{ \large \bf Tables \rm}
\begin{table}[h]
\caption{Parameters of resonances in the fit to the
$\pi^{+}\pi^{-}\pi^{+}\pi^{-}$ mass spectrum and waves.}
\label{ta:a}
\vspace{1in}
\begin{center}
\begin{tabular}{|c|c|c|c|c|} \hline
 & & & & \\
 &Mass (MeV) & Width (MeV) &Observed & $I(J^{PC})$\\
 & & & decay mode & \\
 & & & & \\ \hline
 & & & & \\
$f_{1}(1285)$  &1281 $\pm$ 1 & 24 $\pm$ 3 &$\rho\pi\pi$ &$0(1^{++})$  \\
 & & & & \\ \hline
 & & & & \\
$f_{0}(1300)$ &1290 $\pm$ 15 & 290 $\pm$ 30 & $\rho\pi\pi$ & $0(0^{++})$ \\
 & & & & \\ \hline
 & & & & \\
$f_{0}(1500)$ &1510 $\pm$ 20 & 120 $\pm$ 35 & $\rho\pi\pi$ & $0(0^{++})$ \\
 & & & & \\ \hline
 & & & & \\
$f_{0}(2000)$ &2020 $\pm$ 35 & 410 $\pm$ 50 & $\rho\pi\pi$ & $0(0^{++})$ \\
 & & & & \\ \hline
 & & & &\\
$f_{2}(1900)$ &1960 $\pm$ 30 &460 $\pm$ 40 &$a_{2}(1320)\pi$ &$0(2^{++})$ \\
 & & &$f_{2}(1270)\pi\pi$ &\\
 & & & & \\ \hline
 & & & &\\
$\eta_{2}(1620)$ &1620 $\pm$ 20 &180 $\pm$ 25 &$a_{2}(1320)\pi$ &$0(2^{-+})$ \\
 & & & & \\ \hline
 & & & &\\
$\eta_{2}(1875)$ &1840 $\pm$ 25 &200 $\pm$ 40 &$a_{2}(1320)\pi$ &$0(2^{-+})$ \\
 & & & & \\ \hline
\end{tabular}
\end{center}
\end{table}
\newpage
\begin{table}[h]
\caption{Resonance production as a function of $dP_T$
expressed as a percentage of its total contribution.}
\label{ta:b}
\vspace{1in}
\begin{center}
\begin{tabular}{|c|c|c|c|} \hline
 & & &  \\
 &$dP_T$$\leq$0.2 GeV & 0.2$\leq$$dP_T$$\leq$0.5 GeV &$dP_T$$\geq$0.5 GeV\\
 & & & \\ \hline
 & & & \\
$f_{1}(1285)$  &5.8 $\pm$ 0.4 & 49.3 $\pm$ 1.3 &44.8 $\pm$ 0.8 \\
 & & & \\ \hline
 & & & \\
$f_{0}(1300)$  &28.4 $\pm$ 1.0 & 43.2 $\pm$ 1.0 &28.4 $\pm$ 2.0 \\
 & & & \\ \hline
 & & & \\
$f_{0}(1500)$  &30.8 $\pm$ 1.0 & 50.5 $\pm$ 1.0 &18.8 $\pm$ 2.0 \\
 & & & \\ \hline
 & & & \\
$f_{0}(2000)$  &12.2 $\pm$ 2.0 & 53.1 $\pm$ 2.0 &34.7 $\pm$ 2.5 \\
 & & & \\ \hline
 & & & \\
$f_{2}(1900)$  &31.1 $\pm$ 2.0 & 49.2 $\pm$ 2.0 &19.7 $\pm$ 1.0 \\
 & & & \\ \hline
 & & & \\
$\eta_{2}(1620)$  &2.4 $\pm$ 1.0 & 50.6 $\pm$ 3.4 &46.8 $\pm$ 3.3 \\
 & & & \\ \hline
 & & & \\
$\eta_{2}(1875)$  &0.4 $\pm$ 0.4 & 39.3 $\pm$ 6.0 &60.3 $\pm$ 6.0 \\
 & & & \\ \hline
\end{tabular}
\end{center}
\end{table}
\newpage
{ \large \bf Figures \rm}
\begin{figure}[h]
\caption{The
$\pi^{+}\pi^{-}\pi^{+}\pi^{-}$
effective mass spectrum a) for the total data with fit using
3 Breit-Wigners
b) for $dP_T$~$\leq$~0.2~GeV,
c) for 0.2~$\leq$~$dP_T$~$\leq$~0.5~GeV and
d) for $dP_T$~$\geq$~0.5~GeV.}
\label{fi:a}
\end{figure}
\begin{figure}[h]
\caption{Results of the spin parity analysis.
The superimposed curves are the resonance contributions coming from
the fits described in the text.}
\label{fi:b}
\end{figure}
\begin{figure}[h]
\caption{Results of the spin parity analysis as a function of $dP_T$.
}
\label{fi:c}
\end{figure}
\newpage
\begin{center}
\epsfig{figure=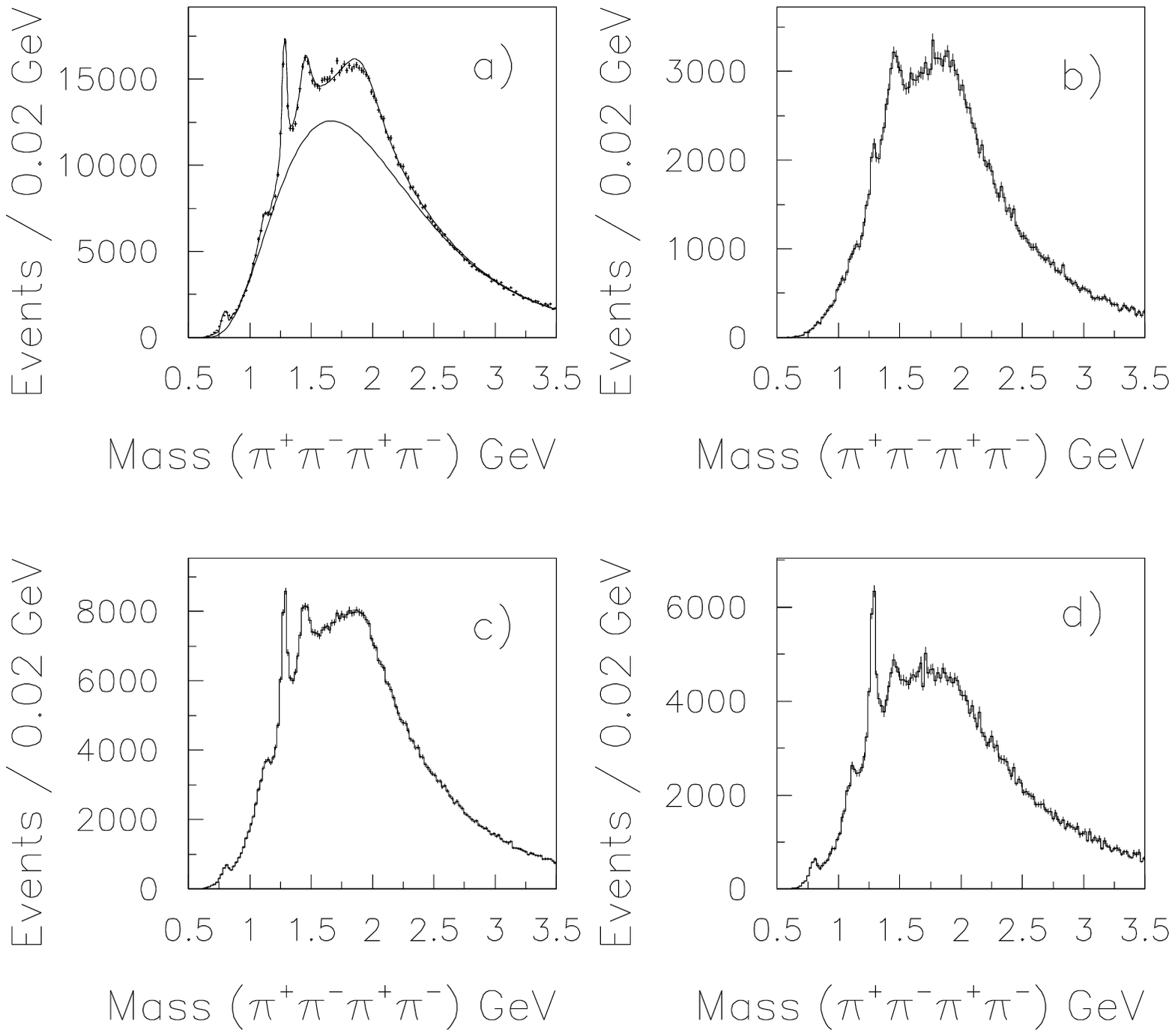,height=22cm,width=17cm}
\end{center}
\begin{center} {Figure 1} \end{center}
\newpage
\begin{center}
\epsfig{figure=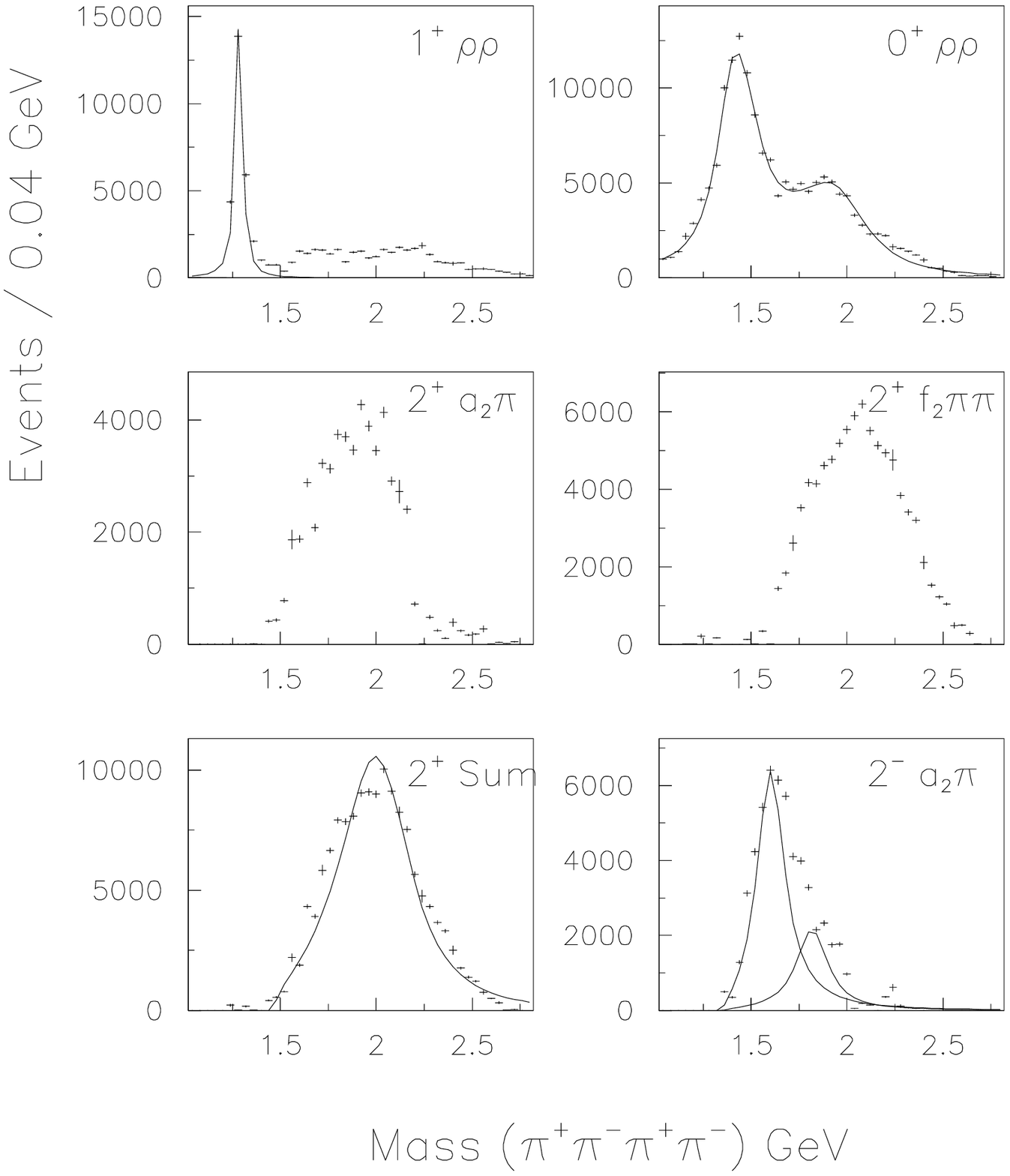,height=22cm,width=17cm}
\end{center}
\begin{center} {Figure 2} \end{center}
\newpage
\begin{center}
\epsfig{figure=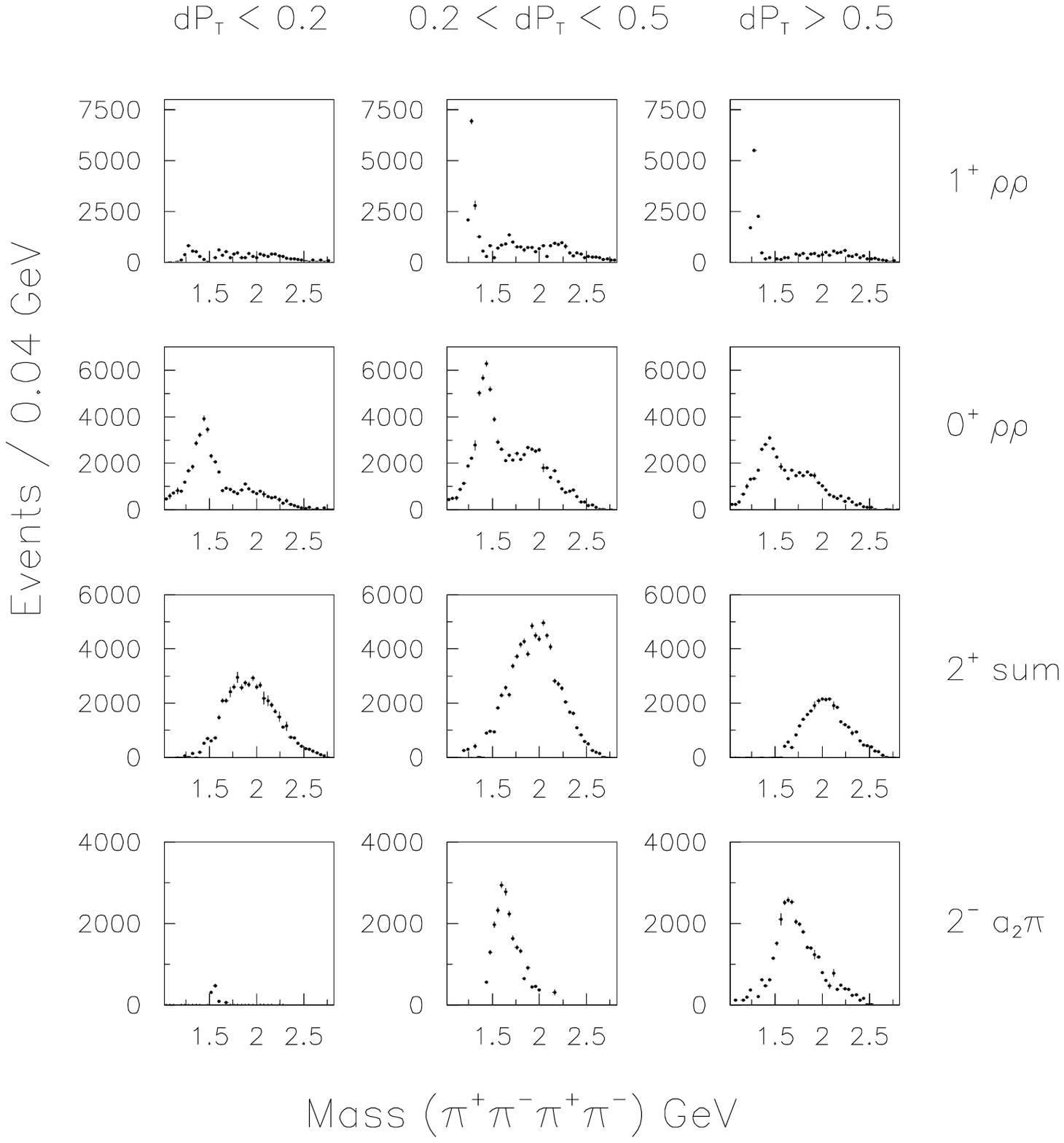,height=22cm,width=17cm}
\end{center}
\begin{center} {Figure 3} \end{center}
\end{document}